\documentclass{article}

\usepackage[final]{nips_2016}

\usepackage[utf8]{inputenc} 
\usepackage[T1]{fontenc}    
\usepackage{hyperref}       
\usepackage{url}            
\usepackage{booktabs}       
\usepackage{amsfonts}       
\usepackage{nicefrac}       
\usepackage{microtype}      
\usepackage{graphicx}
\usepackage{amsmath}
\usepackage{chngcntr}
\usepackage{multirow}

\usepackage{xr-hyper} 
\usepackage{hyperref} 

\newcommand{\graphset}{\mathcal G}
\newcommand{\nodeset}{\mathcal V}
\newcommand{\edgeset}{{\mathcal E}}
\newcommand{\covariancematrix}{\Sigma}
\newcommand{\precisionmatrix}{{\Sigma^{\text -1}}}

\newcommand{\edge}{e}

\newcommand{\rnp}{\mathbb{R}^{n\times p}}
\newcommand{\rpp}{\mathbb{R}^{p\times p}}

\newcommand{\E}{\mathrm{E}}
\newcommand{\Var}{\mathrm{Var}}

\newcommand{\argmin}{\mathop{\mathrm{argmin}}}
\newcommand{\argmax}{\mathop{\mathrm{argmax}}}

\title{Generalized Stability Approach for Regularized Graphical Models}

 \author{
   Christian L.~M\"uller\\
   Simons Center for Data Analysis\\
  Simons Foundation\\
  New York, NY 10010 \\
   \texttt{cmueller@simonsfoundation.org} \\
  \And
  Richard Bonneau \\
  Simons Center for Data Analysis and\\
  Departments of Computer Science and Biology\\
   New York University\\
   New York, NY 10012 \\
 \texttt{rb133@nyu.edu} \\
 \AND
     Zachary D. Kurtz\\   
     Departments of Microbiology and Medicine\\
     New York University School of Medicine\\
     New York, NY 10016 \\
     \texttt{zachary.kurtz@med.nyu.edu} \\
}

\begin{document}

\maketitle

\begin{abstract}
Selecting regularization parameters in penalized high-dimensional graphical models in a principled, data-driven, and computationally efficient manner continues to be one of the key challenges in high-dimensional statistics. We present substantial computational gains and conceptual generalizations of the Stability Approach to Regularization Selection (StARS), a state-of-the-art graphical model selection scheme. Using properties of the Poisson-Binomial distribution and convex non-asymptotic distributional modeling we propose lower and upper bounds on the StARS graph regularization path which results in greatly reduced computational cost without compromising regularization selection. We also generalize the StARS criterion from single edge to induced subgraph (graphlet) stability. We show that simultaneously requiring edge and graphlet stability leads to superior graph recovery performance independent of graph topology. These novel insights render Gaussian graphical model selection a routine task on standard multi-core computers.  
\end{abstract}
\section{Introduction}
Probabilistic graphical models \citep{Lauritzen:1996} have become an important scientific tool for finding and describing patterns in high-dimensional data. 
Learning a graphical model from data requires a simultaneous estimation of the graph  and of the probability distribution that factorizes according to this graph. In the Gaussian case, the underlying  graph is determined by the non-zero entries of the precision matrix (the inverse of the population covariance matrix). Gaussian graphical models have become popular after the advent of computationally tractable estimators, such as neighborhood selection \citep{Meinshausen2006} and sparse inverse covariance estimation \citep{Banerjee2008,Yuan:2007}. State-of-the-art solvers are the Graphical Lasso (\texttt{GLASSO}) \citep{Friedman2008} and the QUadratic approximation for sparse Inverse Covariance estimation (\texttt{QUIC}) method \citep{Hsieh2014}. 

Any neighborhood selection and inverse covariance estimation method requires a careful calibration of a regularization parameter because the actual model complexity is not known \textit{a priori}. 
State-of-the-art tuning parameter calibration schemes include cross-validation, (extended) information criteria (IC) such as Akaike IC and Bayesian IC \citep{Yuan:2007,Foygel2010}, and the Stability Approach to Regularization Selection (StARS) \citep{Liu2010}. The StARS method is particularly appealing because it shows superior empirical performance on synthetic and real-world test cases \citep{Liu2010} and has a clear interpretation: StARS seeks the minimum amount of regularization that results in a sparse graph whose edge set is reproducible under random subsampling of the data at a fixed proportion $1-\beta$ with standard setting $\beta=0.1$ \citep{Zhao2012a}. Regularization parameter selection is thus determined by the concept of stability rather than regularization strength. However, two major shortcomings in StARS are computational cost and optimal setting of $\beta$. StARS must repeatedly solve costly global optimization problems (neighborhood or sparse inverse covariance selection) over the entire regularization path for $N$ sets of subsamples (where the choice of $N$ is user-defined). Also, there may be no universally optimal setting of $\beta$ as edge stability is strongly influenced by the underlying unknown topology of the graph \citep{Ravikumar2011b}.

In this paper we develop an approach to both of these shortcomings in StARS. We speed up StARS by proposing $\beta$-dependent lower and upper bounds $\lambda_\mathrm{lb}, \lambda_\mathrm{ub}$ on the regularization path from as few as $N=2$ subsamples such that $\lambda_\beta \in [\lambda_\mathrm{lb}, \lambda_\mathrm{ub}]$ with high probability (Sec.~\ref{sec:stars}). This particularly implies that the lower part of regularization path (resulting in dense graphs and hence computationally expensive optimization) does not need to be explored for future samples without compromising selection quality. Secondly, we generalize the concept of edge stability to induced subgraph (graphlet) stability. Building on a recently introduced graphlet-based graph comparison scheme \citep{Yaveroglu2014} we introduce a novel measure that accurately captures variability of small induced subgraphs across graph estimates (Sec.~\ref{sec:GStARS}). We show that simultaneously requiring edge and graphlet stability leads to superior regularization parameter selection on realistic synthetic benchmarks (Sec.~\ref{sec:benchmark}). To showcase real-world applicability we infer, in Sec.~\ref{americangut}, the largest-to-date gut microbial ecological association network from environmental sequencing data at dramatic speedup. All proposed methods, along with several efficient parallel software utilities for shared-memory and cluster environments, are implemented in R and \texttt{MATLAB} and will be made freely available at the authors' github repository.


\section{Gaussian graphical model inference}
\label{sec:ggm}
We consider $n$ samples from a $p$-dimensional Gaussian
distribution $\mathcal N (0,\Sigma)$ with positive definite, symmetric covariance matrix
$\covariancematrix\in\rpp$ and symmetric precision matrix $\Theta = \precisionmatrix$.
The samples are summarized in the matrix $X\in\rnp$ where $X_{ij}$ corresponds to the
$j$th component of the $i$th sample. The Gaussian distribution $\mathcal N (0,\Sigma)$
can be associated with an undirected graph $\graphset=(\nodeset,\edgeset)$, where 
$\nodeset=\{1,\dots,p\}$ is the set of nodes and $\edgeset=\nodeset\times \nodeset$ the
set of (undirected) edges that consists of all pairs
$(i,j),(j,i)\in\nodeset\times\nodeset$ that fulfill  $i\neq j$ and  $(\Theta)_{ij}\neq 0$.
We denote by $\edge_{ij}$ (or $\edge_{ji}$) the edge that corresponds to the pair $(i,j),(j,i)$ and by $E:=|\edgeset|=\Vert \Theta \Vert_{0}$ the number of edges in the graph~$\graphset$. An alternative single indexing $l=1,\ldots,L$ with $L = \frac{1}{2}p(p-1)$ of all edges follows the column-wise order of the lower triangular part of the adjacency matrix of $\graphset$, i.e., edge $(1,1) \rightarrow 1, (2,1) \rightarrow 2, \ldots, (3,2) \rightarrow p+1,\ldots, (p,p-1) \rightarrow L$.   

\subsection{Sparse inverse covariance estimation}
One popular way to estimate the non-zero entries of the precision matrix $\Theta$ from data~$X$, or equivalently, the set of weighted edges~$\edgeset$ from~$X$, relies on minimizing the negative penalized log-likelihood. In the standard Gaussian setting, the estimator with positive regularization parameter $\lambda$ reads:
\begin{equation}
\label{eq:spic}
\hat{\Theta}(\lambda) =  \underset{ \Theta \succ 0}{\arg\min} \left( -\log\det(\Theta)+\mathrm{tr}(\Theta \hat \Sigma)+\lambda\left\Vert \Theta \right\Vert _{1} \right) \, ,
\end{equation}
where 
$\Theta \succ 0$ denotes the set of real positive definite matrices, $\hat \Sigma$ the sample covariance estimate, $\| \cdot \|_1$ the element-wise L1 norm, and $\lambda \ge 0$ a scalar tuning parameter. For $\lambda=0$, the expression is identical to the maximum likelihood estimate of a normal distribution $\mathcal{N}(x | 0,\Sigma)$. For non-zero $\lambda$, the objective function encourages sparsity of the underlying precision matrix $\hat \Theta(\lambda)$ (and graph $\hat \graphset(\lambda)$, respectively). This estimator was shown to have theoretical guarantees on consistency and recovery under normality assumptions. Recent theoretical work \citep{Ravikumar2011b} also shows that distributional assumptions can be considerably relaxed, and that the estimator is applicable to a larger class of problems, including inference on discrete (count) data or on data transformed by nonparametric approaches.

\subsection{Efficient optimization algorithms}
A popular first-order method for solving Eq.~(\ref{eq:spic}) is the Graphical Lasso (\texttt{GLASSO}) \citep{Friedman2008} which solves the row sub-problem of the dual of Eq.~(\ref{eq:spic}) using coordinate descent. The arguably fastest method to date is the QUadratic approximation for sparse Inverse Covariance estimation (QUIC) \citep{Hsieh2014} which iteratively applies Newton's method to a quadratic approximation of Eq.~(\ref{eq:spic}). The key features in \texttt{QUIC} are (i) efficient computation of
the Newton direction ($O(p)$ instead of $O(p^2)$) by exploiting the special structure of
the Hessian in the approximation and (ii) automatic on-the-fly partitioning of variables
into a \textit{fixed} and a \textit{free} set. Newton updates need to be applied only to
the set of free variables which can dramatically reduce the run time when the estimated graph is sparse. Importantly, this strategy generalizes the observations made by \cite{Witten2011} and \cite{Mazumder2012} that inverse covariance estimation can be considerably sped up when the underlying matrix has block-diagonal structure, which can be easily identified by thresholding the absolute values of $\hat \Sigma$. \cite{Hsieh2014}'s large-scale performance comparison of all state-of-the-art algorithms reveals that (i) \texttt{QUIC} and, to a lesser extent, \texttt{GLASSO} are the only methods that efficiently solve large-scale graphical models and (ii) both methods show a dramatic increase in run time when the regularization parameter $\lambda$ is small (resulting in dense graph estimates; see \cite{Hsieh2014}, Fig.7). Thus, finding a lower bound on $\lambda$ that does not interfere with model selection quality is highly desirable when learning graphical models with \texttt{QUIC} and \texttt{GLASSO}.    
 
\section{Stability-based graphical model selection}
\label{sec:stars}
Stability-based model selection schemes have recently gained considerable attention due to their theoretical and practical appeal \citep{Meinshausen:2010}. The Stability Approach to Regularization Selection (StARS) \citep{Liu2010} shows particular promise for graphical model selection and is the primary application developed below.  

\begin{figure}[h]
   \centering
   \includegraphics[width=0.79\linewidth]{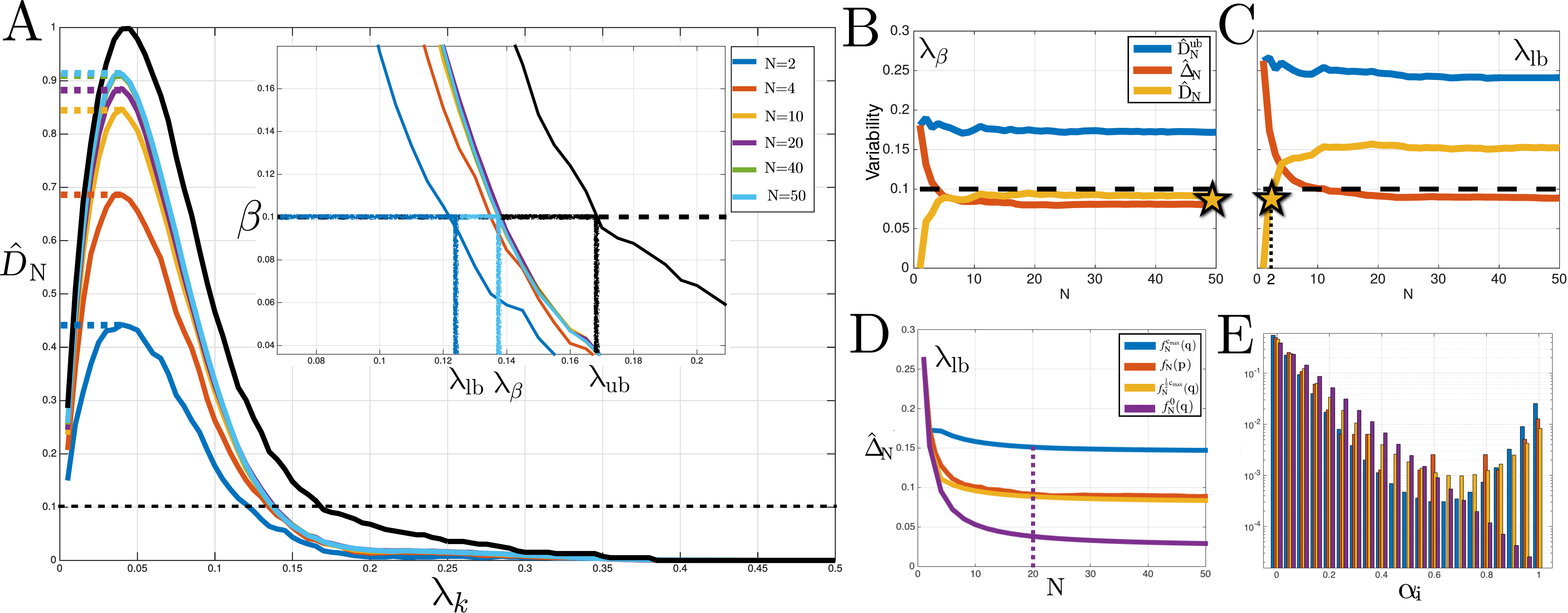}
\caption{A) Typical behavior of $\hat{D}_{N}(\lambda_k)$ with increasing $N$ on the neighborhood graph example  \citep{Liu2010} with $n=800$, $p=40$. The colored dashed lines on all curves show the monotonized $\bar{D}_{N}$. The black dashed line shows the variability threshold $\beta=0.1$. 
The $\lambda_\mathrm{k}$ selected at $N=2$ ($\lambda_\mathrm{lb}$) is a lower bound on $\lambda_\beta$. The black upper bound curve ($\lambda_\mathrm{ub}$) arises from a binomial approximation with average edge probability (see Sec.~\ref{subsec:pbd}) \label{fig-totvar}. B) Total variability $\hat{D}_{N}(\lambda_k)$ (orange curve) and its decomposition into $\hat{D}^\mathrm{ub}_{N}(\lambda_k)$ (blue curve) and $\Delta_{N}(\lambda_k)$ (red curve) for the nearest neighbor graph example at $\lambda_\beta$ for $N\le50$; C) Same as B) for $\lambda_\mathrm{lb}$.  \label{fig-totvardecomp} D) Within-probability variability  $\hat{\Delta}_{N}(\lambda_\mathrm{lb})$ for the nearest neighbor graph example ($L=780, \epsilon = 3/L$) (red curve), corresponding maximum entropy models $f^{0}_\mathrm{N}(\mathbf{q})$ (purple curve), $f^{\mathrm{c}_\mathrm{max}}_\mathrm{N}(\mathbf{q})$ (blue curve) with $c_\mathrm{max}=1/2$, and best approximation (orange curve, $c=1/2 \, c_\mathrm{max}=1/4$) to $\hat{\Delta}_{N}(\lambda_\mathrm{lb})$ for $N\le50$. E) Discrete probability distributions $f_\mathrm{N}$ for $N=20$ subsamples (color scheme as in D). \label{fig-var}}
\end{figure}
\subsection{StARS revisited}
\label{subsec:stars}
StARS assesses graph stability via subsampling. 
Let $b(n)$ be the size of a subsample with $1<b(n)<n$ with $b(n)=\lfloor10\sqrt{n}\rfloor$ the recommended choice. We draw $N$ random subsamples $S_{1},\ldots,S_{r},\ldots,S_{N}$ of size $b(n)$ and solve Eq.~(\ref{eq:spic}) for each $S_{r}$ over a grid of positive regularization parameters $\mathcal{P}=\left\{ \lambda_{1},\ldots,\lambda_{k},\ldots,\lambda_{K}\right\}$ with $\lambda_{k}<\lambda_{k+1}$. 
We denote the estimated precision matrices and graphs by $\hat \Theta(\lambda_k)$ and $\hat \graphset(\lambda_k) = (\nodeset(\lambda_k), \edgeset(\lambda_k))$. For each node pair $(i,j) \in \edgeset_r(\lambda_{k})$ let the indicator $\psi^{\lambda_k}_{ij}(S_{r})=1$ if the algorithm infers an edge for the given subsample $S_r$, and $\psi^{\lambda_k}_{ij}(S_{r})=0$ otherwise. StARS estimates the probability $\theta^{\lambda_k}_{ij}=\mathbb{P}\left(\psi^{\lambda_k}_{ij}(S_{r})=1\right)$ via the U-statistic of order $b(n)$ over $N$ subsamples
$\hat{\theta}^{\lambda_k}_{ij}=\frac{1}{N}\underset{r=1}{\overset{N}{\sum}}\psi^{\lambda_k}_{ij}(S_{r}) \, .
$
Using the estimate $\hat{\xi}_{ij}^{\lambda_k}=2\hat{\theta}_{ij}^{\lambda_k}\left(1-\hat{\theta}_{ij}^{\lambda_k}\right)$, i.e., twice the variance of the Bernoulli indicator of the edge $e_{ij}$ across $N$ subsamples. \cite{Liu2010} defines the total instability (or variability) of all edges as:
%
$
\hat{D}_{N}(\lambda_k)=\frac{2\sum_{i>j}\hat{\xi}_{ij}^{\lambda_k}}{p(p-1)/2} \,.
$
%
\cite{Liu2010}  propose to monotonize $\hat{D}_{N}$ via $\bar{D}_{N}(\lambda)=\underset{\lambda_K\ge\lambda\ge0} \sup \hat{D}_{N}(\lambda)$ and show that the selection procedure $\lambda_{\beta}=\inf\left\{ \bar{D}_{N}(\lambda) \leq \beta \colon \lambda \in \mathcal{P} \right\}$ guarantees asymptotic partial sparsistency for a fixed threshold $\beta$ (see Thm.~2, \cite{Liu2010}). 
This means that the true graph $\graphset$ is likely to be included in the edge set of $\hat{\graphset}(\lambda_\beta)$ with few false negatives with standard setting $\beta = 0.1$.  
No formal guidance is given in \citet{Liu2010} regarding the number $N$ of subsamples despite the fact that this has a great influence on the run time when applying StARS in a sequential setting (the default in the \texttt{huge} implementation \citep{Zhao2012a} is $N=20$).
\vspace{-2mm}
\paragraph*{Observation 1:} \label{obs:1} For all synthetic examples given in \citep{Liu2010}, $N\ge20$ produces smooth variability curves $\hat{D}_{N}$ that lead to accurate selection of $\lambda_\beta$. For small $N$, the variability $\hat{D}_{N}$ is uniformly underestimated over the regularization path $\mathcal{P}$.   
Figure~\ref{fig-totvar} shows the typical behavior of $\hat{D}_{N}(\lambda_k)$ with increasing $N$ on the nearest-neighbor graph construction used in \cite{Liu2010} as synthetic test case (here $n=800$, $p=40$, $E=66$) where $\hat{D}_{N}(\lambda_k)$ estimated for $N=2$ provides a uniform lower bound on the variability curve for large $N$.
\vspace{-2mm}
\paragraph*{Observation 2:} For fixed $\lambda_k$ the total edge variability $\hat{D}_{N}(\lambda_k)$ can be interpreted as normalized variance of the sum of $L$ independent Bernoulli random variables with $L$ distinct success probabilities $p_1,\ldots,p_\mathrm{L}$, one for each potential edge in the graph. The sum of $L$ independent Bernoulli variables follows a Poisson Binomial distribution over $\{0,1,\ldots,L\}$.
\subsection{Modeling total variability with the Poisson Binomial distribution}
\label{subsec:pbd}
Denote by $Y_L$ the sum of $L$ Bernoulli indicators with individual trial probabilities $\mathbf{p} = (p_1,\ldots,p_l,\ldots,p_L)$. Then $Y_L$ follows a Poisson Binomial
distribution (PBD) with probability mass function $f_\mathrm{PB}(y;\mathbf{p})= \underset{A \in \mathcal{F}_y} \sum \left(\underset{l \in A} \prod p_l\right)\left(\underset{k \in A^c}\prod (1-p_k)\right)$ 
with $\mathcal{F}_y = \{A: A \subseteq {1,\ldots,L}, |A| = y\}$, expectation $\E(Y_L) = \sum_{l=1}^{L} p_l$, and variance $\Var(Y_L) = \sum_{l=1}^{L} p_l (1-p_l)$ \citep{Poisson1837}. A well-known fact about $\Var(Y_L)$ is its decomposition \citep{Wang1993} 
%
$
\Var(Y_L) = L \bar{p}(1-\bar{p}) - L \sigma_p^2 \, ,
$
%
where $\bar{p}=\frac{1}{L}\sum_{l=1}^L p_l$ is the mean probability and $\sigma_p^2 = \sum_{l=1}^L (p_l-\bar{p})^2$ is the variance within the probability vector $\mathbf{p}$. Denote by $\bar{Y}_L= Y_L/L$ with $\E(\bar{Y}_L) = \bar{p}$. This implies that the $\Var(\bar{Y}_L) \leq \bar{p}(1-\bar{p})/L$, and the upper bound is attained when all $p_i$ are homogeneous. By applying Chebychev's inequality we get for any $\epsilon$ and $\mathbf{p}$ (see, e.g.,\cite{Wang1993}, Corr.~1):
%
$
\mathbb{P}\left(|\bar{Y}_L - \bar{p}| > \epsilon \right) \leq 1/(4L\epsilon^2) \, .
$
%
\subsubsection{An upper bound on total variability}
\label{subsec:ub}
Recall that StARS estimates the probabilities $\theta^{\lambda_k}_{ij}$ for each node pair $(i,j)$ via subsampling. Using the single-index notation we see that the quantities $\{ \hat \theta^{\lambda_k}_{l}: l=1,\ldots,L \}$ are $L$ approximately independent, non-identical probability estimates $\hat p_1,\ldots,\hat p_\mathrm{L}$ for each edge $e_l$. For any fixed $\lambda_k$ the StARS variability $\hat{D}_{N}(\lambda_k)$
is thus a scaled estimator for the variance of the PBD associated with $L=p(p-1)/2$ edge probabilities. 
This observation has immediate consequences for the non-asymptotic behavior of StARS total variability at small $N$. 
The Chebychev bound on $\bar p$ shows that we get an $O(1/\sqrt{L})$ approximation to $\bar p$ with high probability, implying that the average edge probability is extremely accurate for all relevant graph sizes even for a small number of subsamples $N$. This leads to the following practical upper bound on the StARS variability. Let $\bar \theta^{\lambda_k} = \frac{1}{L} \sum_{l=1}^L \hat \theta^{\lambda_k}_{l}$ be the average edge probability estimate. Then the quantity   
%
$
\hat{D}^\mathrm{ub}_{N}(\lambda_k) = 4 \bar \theta^{\lambda_k} (1-\bar \theta^{\lambda_k}) \,
$
%
with $\hat{D}^\mathrm{ub}_{N}(\lambda_k) \in [0,1]$ is an upper bound on $\hat{D}_{N}(\lambda_k)$  with high probability. Using the monotonized version $\bar{D}^\mathrm{ub}_{N}$ we thus define $\lambda_\mathrm{ub}=\inf\left\{ \bar{D}^\mathrm{ub}_{N}(\lambda) \leq \beta \colon \lambda \in \mathcal{P} \right\}$ for user-defined $\beta$. Figure~\ref{fig-totvar}A) shows the typical behavior of $\hat{D}^\mathrm{ub}_{2}$ over $\mathcal{P}$ (black curve) for the neighborhood graph example \citep{Liu2010}. Estimates of $\hat{D}^\mathrm{ub}_{N}(\lambda_k)$ for the same graph example at two different $\lambda_k$ across $N$ are shown in Fig.~\ref{fig-totvardecomp}B),C) (blue curves).
%
%
\subsubsection{Maximum entropy bounds on $\hat{\Delta}_{N}(\lambda_k)$}
\label{subsec:lb}
The variance decomposition for the PBD 
together with the bounds on $\bar p$, also imply that \textbf{Observation 1} can only stem from an \textit{overestimation} of $\sigma_p^2$ at small $N$. In StARS notation the quantity  
%
$
\hat{\Delta}_{N}(\lambda_k) = 4 \sum_{l=1}^L (\theta_l^{\lambda_k} -\bar \theta^{\lambda_k})^2 \, ,
$
%
is a scaled version of the variance and is referred to as the within-probability variability $\hat{\Delta}_{N}(\lambda_k) \in [0,1]$ with $\hat{D}_{N}(\lambda_k) = \hat{D}^\mathrm{ub}_{N}(\lambda_k)-\hat{\Delta}_{N}(\lambda_k)$.
Figure~\ref{fig-totvardecomp}B,C show the monotonic decrease of $\hat{\Delta}_{N}$ (red curves) at two different $\lambda_k$. The typical sharp drop in $\hat{\Delta}_{N}(\lambda_k)$ at small $N$ suggests that we can define the following lower bound on the regularization path $\mathcal{P}$: $\lambda_\mathrm{lb}=\inf\left\{ \bar{D}_{2}(\lambda) \leq \beta \colon \lambda \in \mathcal{P} \right\}$ for user-defined $\beta$. However, this lower bound is only valid if $\hat{\Delta}_{N}(\lambda_k)$ shows sufficient decrease for $N>2$. While this non-asymptotic behavior cannot be true for all possible edge probability distributions, we here introduce  graph-independent discrete, non-parametric models for the distribution of edge probabilities emerging at relevant $\beta$ values (e.g., $\beta = 0.1$) using the maximum entropy principle and convex optimization under prior constraints (\cite{Boyd2003}, Chap.~7). Denote by $f_N(\mathbf{q})$ the distribution of edge probabilities with mean $\bar q$ and variance $\sigma_\mathbf{q}$ after $N$ subsamples.  $f_N(\mathbf{q})$ is a discrete distribution over $N+1$ distinct locations $\{\alpha_0 = 0, \alpha_1=1/N,\ldots,\alpha_i = i/N,\ldots,\alpha_N=1\}$ and edge probabilities $\mathbb{P}(Z=\alpha_i)=q_i$ (in our case $Z$ represents edges $e_l$). Note that the probability simplex $\mathcal{S} = \{\mathbf{q} \in \mathbb{R}^{N+1} | q_i \in [0,1], \sum_{i=0}^{N}q_i =1 \}$ comprises all possible probability distributions for the random variable $Z$ taking values at $\{\alpha_0,\ldots,\alpha_i,\ldots,\alpha_N\}$. We estimate maximum entropy models of $f_N(\mathbf{q})$ for any finite $N>2$ under the following prior knowledge from $N=2$: (i) $\bar q$ is known up to $\epsilon$, (ii) $f_2(\mathbf{q})$ is right-skewed with most mass at $\alpha_0$, approximately $\beta$ percent at $\alpha_1=1/2$, and small mass at $\alpha_2=1$, (iii) $\forall N$ $q_0$ and $q_N$ are upper bounded by their observed values $\hat q_0$ and $\hat q_2$ at $N=2$, and (iv) the empirical distribution for $N>2$ is bimodal with modes at $\alpha_0$ (no edge) and $\alpha_N$ (high probability edge). This prior knowledge can be formulated as convex constraints in the following convex program $\mathcal{C}(\mathbf{q},c)$: 
\begin{equation}
 f^\mathrm{c}_\mathrm{N}(\mathbf{q})=\argmax_{\mathbf{q} \in \mathcal{S}}   \left (-\sum_{i=0}^{N} q_i \log{q_i}  \right) \,
  \mathrm{s.t.}\,  | \sum_{i=0}^N \alpha_i q_i - \bar q|  \le \epsilon \, ,   
      q_0 \le \hat{q}_0, \, q_N \le \hat{q}_2 \, , 
      \sum_{i=0}^N (\alpha_i^3 - c \alpha_i) q_i \ge 0 \, . 
    \label{con:3}  
\end{equation}
For any prediction $f_N(\mathbf{q})$ for $N>2$, the program $\mathcal{C}(\mathbf{q},c)$ requires as data input the empirical estimates $\bar q$, $\epsilon$, $\hat{q}_0$, and $\hat{q}_2$ from StARS for $N=2$ subsamples. The parameter $c \in [0, c_\mathrm{max}]$ in the last constraint of Eq.~(\ref{con:3}) controls the bimodality of the maximum entropy distribution. The setting $c=0$ results in trivial non-negative skewness constraint and thus an exponential distribution. The scalar $c_\mathrm{max}$ is the largest $c$ that still leads to feasibility of $\mathcal{C}(\mathbf{q},c)$ while promoting distributions with higher probability mass near $\alpha_N$.   
\vspace{-3mm}
\paragraph*{Observation 3:} All maximum entropy distributions consistent with the convex program $\mathcal{C}(\mathbf{q},c)$ show monotonic decrease of $\sigma_\mathbf{q}$ with increasing $N$. The variances $\sigma_\mathbf{p}$ of all distributions from StARS at relevant $\beta$ are bounded by the variances of the distributions $f^\mathrm{0}_\mathrm{N}(\mathbf{q})$ and $f^{\mathrm{c}_\mathrm{max}}_\mathrm{N}(\mathbf{q})$ for all $N$.  

%
%
Figure~\ref{fig-var}D shows the monotonic decrease of $\hat{\Delta}_{N}(\lambda_\mathrm{lb})$ for three maximum entropy models $c = \{0,1/2c_\mathrm{max}, c_\mathrm{max}\}$ with $c_\mathrm{max}=1/2$ consistent with prior knowledge from the nearest neighbor graph example, and Fig.~\ref{fig-var}E the discrete probability distributions $f_\mathrm{N}$ at $N=20$ (the standard setting in StARS), respectively. The generality of this behavior is strongly supported by empirical results across different graph classes, dimensions, and samples sizes (see Sec.\ref{sec:benchmark}).

\subsection{Bounded StARS (B-StARS)}
Modeling StARS variability with the Poisson Binomial Distribution along with maximum entropy predictive bounds from convex optimization provides strong evidence that $N=2$ subsamples suffice to get lower and upper bounds on the regularization path $\mathcal{P}$ that contain $\lambda_\beta$ for $N>2$ with high probability whenever the graph is sufficiently large and sparse. This suggests the following Bounded StARS (B-StARS) approach:
\begin{enumerate}
\item Solve Eq.~\ref{eq:spic} using two subsamples and record $\bar{D}_{2}$ over the entire $\mathcal{P}$. Set target $\beta$. 
\item Compute $\lambda_\mathrm{lb}=\inf\left\{ \bar{D}_{2}(\lambda) \leq \beta \colon \lambda \in \mathcal{P} \right\}$ and
$\lambda_\mathrm{ub}=\inf\left\{ \bar{D}^\mathrm{ub}_{2}(\lambda) \leq \beta \colon \lambda \in \mathcal{P} \right\}$
\item Set $\mathcal{P}(\beta)=[\lambda_\mathrm{lb},\lambda_\mathrm{ub}] \subset \mathcal{P}$ and record $\mathrm{gap}_\mathrm{b} = (\lambda_\mathrm{ub}-\lambda_\mathrm{lb})$. If $\mathrm{gap}_\mathrm{b} > 0$ solve Eq.~\ref{eq:spic} over $\mathcal{P}(\beta)$ for $N-2$ subsamples. Record $\mathrm{gap}_\beta = (\lambda_\beta-\lambda_\mathrm{lb})$. If $\mathrm{gap}_\beta >0$ B-StARS is equivalent to StARS. \end{enumerate}
B-StARS leads to a substantial computational speed-up (see Sec.~\ref{subsec:speedup}) and a natural $\beta$-dependent regularization interval $\mathcal{P}(\beta)$ that defines a potentially informative collection of sparse graphs ``neighboring" the actual graph selected by StARS. Statistics on values of $\mathrm{gap}_\beta$ and $\mathrm{gap}_\mathrm{b}$ across a wide range of synthetic test cases are shown in Fig.~\ref{fig:gap}.
\section{Generalized StARS using graphlets}
\label{sec:GStARS}
Combining the variability of individual edges across a collection of $N$ graph estimates $\hat \graphset_r$ into the single scalar quantity $\bar{D}_{N}$ is arguably the simplest and most interpretable way of assessing graph stability. This approach neglects, however, potentially valuable information about coherent stability of a collection of edges, i.e., local subgraphs, or global graph features (e.g., node degree distributions) that is present across different graph estimates over the regularization path $\mathcal{P}$. We show that capturing such information provides additional guidance about regularization parameter selection by extending the StARS concept of edge stability to graphlet stability.
\subsection{Graphlets and the Graphlet Correlation Matrix}
Given an undirected graph $\graphset=(\nodeset,\edgeset)$ and any subset of vertices $\mathcal{V}_{g} \subseteq \nodeset$, the edge set $\mathcal{E}_{g} \subseteq \edgeset$ consisting of all vertex pairs $\{(i,j) | i,j \in \mathcal{V}_{g} \cap (i,j) \in \edgeset \}$ is called the induced subgraph or graphlet $G$ associated with $\mathcal{V}_{g}$. 
\begin{figure}[h]
\begin{center}
\includegraphics[width=0.59\textwidth]{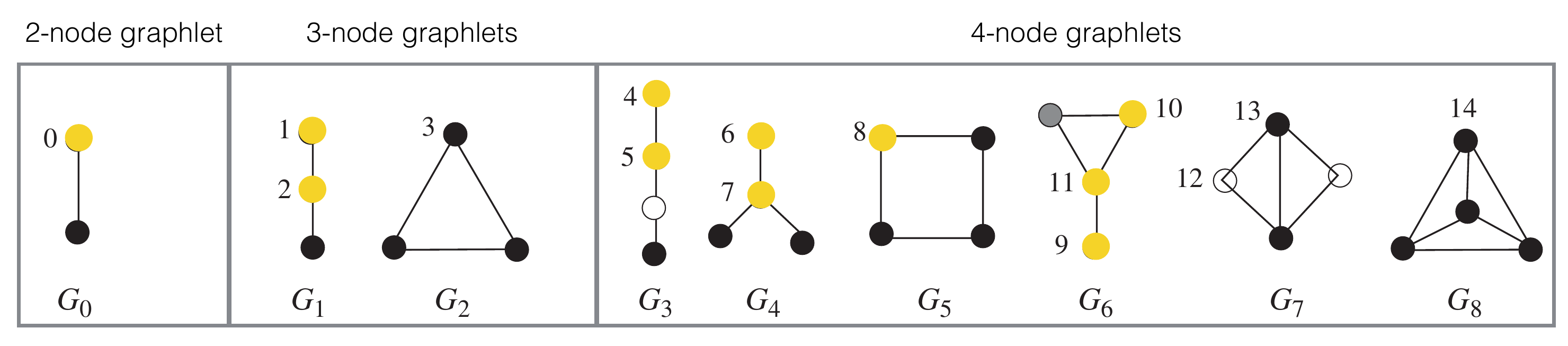}
\end{center}
\caption{All graphlets $G_i,i=0,\ldots,8$ of vertex size $\le 4$. The internal vertex numbering refers to all unique 15 orbits. The 11 yellow nodes represent a set of non-redundant orbits used in the Graphlet Correlation Distance (GCD) (adapted from \cite{Przulj2007}).\label{fig:graphlets}} 
\end{figure}

Graphlets have been extensively used in network characterization and comparison \citep{Przulj2004}. A common strategy is to count the number of graphlets up to a certain size present in a given graph and derive low-dimensional  \textit{sufficient} summary statistics from these counts \citep{Shervashidze2009}. A popular statistics are Graphlet Degree Vectors (GDVs) \citep{Guerrero2008} which provide a vertex-centric graph summary statistic by counting the number of times each vertex is touched by an automorphism group of a graphlet. Considering the collection of all nine graphlets up to size four $G_i,i=0,\ldots,8$ (shown in Fig.~\ref{fig:graphlets}) there exist 15 automorphism groups (or orbits) (numbered $0,\ldots,14$ in Fig.~\ref{fig:graphlets}). Two nodes belong to the same orbit if there exists a bijection of nodes that preserves adjacency (an isomorphic projection). GDVs generalize the classical concept of vertex degree distributions because the degree of a vertex is simply the orbit participation for 2-node graphlets (the first orbit labeled 0 in Fig.~\ref{fig:graphlets}). 
\cite{Yaveroglu2014} established that the information contained in the 15 orbits comprises redundancy that can be reduced to a set of 11 non-redundant orbits (shown in yellow in Fig.~\ref{fig:graphlets}). Each graph $\graphset$ with $p$ vertices can thus be summarized by the Graphlet Degree Matrix (GDM) $M \in \mathbb{N}_0^{p \times 11}$. 
%
While the GDM $M$ has been shown to be an excellent description for graph summarization and comparison \citep{Guerrero2008}, \cite{Yaveroglu2014} demonstrated that the Spearman rank correlation matrix $R = \mathrm{corr}_\mathrm{S}(M) \in \mathbb{R}^{11 \times 11}$ of non-redundant orbits (the ``Graphlet Correlation Matrix") provides sufficient statistical power to identify unique topological signatures in graphs and, hence, to discriminate different graph classes. 
Due to the symmetry in $R$ the description of any graph $\graphset$ can be further compressed by using the lower triangular part of $R$ and storing the entries in column-wise order in the Graphlet Correlation Vector (GCV) $\mathbf{\rho} \in \mathbb{R}^{55}$. 

\subsection{Graphlet stability and model selection}
Given the standard StARS protocol we estimate a collection of $N$ graphs $\hat \graphset_r(\lambda_k) = (\nodeset, \edgeset_r(\lambda_k)), r=1,\ldots,N$ for each $\lambda_k \in \mathcal{P}$. To assess the global topological variability within the graph estimates at fixed $\lambda_k$ we propose the following measure: Let $\mathbf{\rho}^{(r)}$ be the GCV derived from $\hat \graphset_r(\lambda_k)$ and let $\mathrm{dist}(\mathbf{\rho}^{(r)},\mathbf{\rho}^{(s)}) = \sqrt{\sum_{j=1}^{55} (\rho^{(r)}_j - \rho^{(s)}_j)^2}$ be the Euclidean distance between $\mathbf{\rho}^{(r)}$ and $\mathbf{\rho}^{(s)}$ (referred to as Graphlet Correlation Distance (GCD) in \cite{Yaveroglu2014}). Then the total graphlet variability measure over $N$ graph estimates is
%
$
\hat{d}_{N}(\lambda_k) = \frac{2}{N (N-1)} \sum_{r>s} \mathrm{dist}(\mathbf{\rho}^{(r)},\mathbf{\rho}^{(s)}) \, ,
$
%
which is the average Euclidean distance among all GCVs at fixed $\lambda_k$. 
Similar to the total edge variability $\hat D_N$ the total graphlet variability $\hat{d}_{N}$ will converge to zero for collections of very sparse and very dense graphs (i.e., at the boundary of $\mathcal{P}$). However, the intermediate behavior will highly dependent on the topology of the underlying true graph, and $\hat{d}_{N}(\lambda)$ will likely be non-monotonic and potentially multi-modal along $\mathcal{P}$. We propose to use the information contained in $\hat{d}_{N}(\lambda)$ in two ways: (i) as exploratory graph learning tool (illustrated in detail in the Appendix) and (ii) as supporting statistics to improve StARS performance. The key idea for the latter proposition is to require simultaneous edge \textit{and} graphlet stability. This is realized in the following Generalized StARS (G-StARS) scheme:
\begin{enumerate}
\item Set user-defined $\beta$ controlling edge stability $\bar{D}_{N}$.
\item Determine $\mathcal{P}(\beta) \subset \mathcal{P}$ using B-StARS procedure.
\item Select $\lambda_\gamma = \underset{\lambda \in \mathcal{P}(\beta)} \argmin \, \hat{d}_{N}(\lambda)$.
\end{enumerate}
This method ensures (i) that the desired edge stability is approximately satisfied while being locally maximally stable with respect to graphlet variability and (ii) that the computational speed-up gains of B-StARS are maintained.

\section{Numerical benchmarks}
\label{sec:benchmark}
To evaluate both speed-up gains and model selection performance of the proposed StARS schemes we closely follow the computational benchmarks outlined in \cite{Hsieh2014} and \cite{Liu2010}, respectively. For graphical model inference we use the \texttt{QUIC} method \citep{Hsieh2014} as the state-of-the-art inference scheme.  %
\subsection{Speed-up using B-StARS}
\label{subsec:speedup}
The first set of experiments illustrates the substantial speed-up gained using B-StARS without compromising the model selection quality of StARS. We consider the Erd\"os-Renyi graph example (from \cite{Hsieh2014}, Sec.~5.1.1) with $p = 4000$ and $n=2000$. 
We report the baseline wall-clock times for this example in the Appendix in Tbl.~A.1, including run times at fixed $\lambda$ and over the path $\mathcal{P}$ of length $K=20$. We observe substantial increase in run time when running \texttt{QUIC} with StARS subsample size $b(n)=447$ (instead of $n=2000$) due to the dramatic increase in edge density for small $\lambda$ (see $\Vert\hat{\Theta}\Vert_{0}$ columns in Appendix Tbl.~A.1). The first row in Tbl.~\ref{table:quicpulsar} summarizes run times for the full StARS procedure for $N=100$ subsamples. In serial mode, standard StARS would require about a week of computation on a standard laptop. B-StARS reduces this cost by a factor of 16 for this example. For completeness, we also report wall-clock times for embarrassingly parallel batch submissions to a standard multi-core multi-processor cluster systems both in standard and B-StARS mode. Even in this setting, the run time using B-StARS can be reduced by a factor of 2, obtaining statistically equivalent results in less than 3 hours.
\begin{table}[t]
  \caption{Run times for StARS and B-StARS (B-S) in serial and batch mode.}
\label{table:quicpulsar} 
 \centering
\begin{tabular}{cccccc}
\toprule
{\footnotesize{}time (s)} & {\footnotesize{}$N$} & {\footnotesize{}serial} & {\footnotesize{}B-S} & {\footnotesize{}batch} & {\footnotesize{}batch B-S}\tabularnewline
\midrule
{\footnotesize{}Erd\"os-Renyi} & {\footnotesize{}$100$} & {\footnotesize{}$\approx600000$} & {\footnotesize{}$38117$} & {\footnotesize{}$16084$} & {\footnotesize{}$9121$}\tabularnewline
\midrule
{\footnotesize{}American Gut} &{\footnotesize{}$200$} & {\footnotesize{}$929165$} & {\footnotesize{}$21908$} & {\footnotesize{}$30680$} & {\footnotesize{}$6325$}\tabularnewline
\bottomrule
\end{tabular}

\end{table}

\subsection{Model selection using G-StARS}
We next evaluate model selection performance of G-StARS. We follow and considerably extend the computational experiments presented in the original StARS paper \citep{Liu2010}. We generate zero mean multivariate normal data using three different graph/precision matrix models: neighborhood (also termed Geometric) graphs, hub graphs, and Erd\"os-Renyi graphs. For the first two models, we follow the matrix generation scheme outlined in \citep{Liu2010}, for Erd\"os-Renyi graphs we generate positive definite precision matrices with entries in $[-1,1]$ and sparsity level $3/p$. In addition to low-dimensional ($n = 800$, $p = 40$) and high-dimensional ($n = 400$, $p = 100$) settings considered in \citep{Liu2010}, we test StARS and G-StARS for the settings ($n = 200$, $p = 200$) and ($n = 100$, $p = 400$) and used standard $\beta=0.1$.

To evaluate overall graph estimation performance, we report mean (and std) $F_{1}$-scores across all experimental settings (over $200$ repetitions) in Fig.~\ref{fig:f1}. The corresponding precision and recall plots can be found in the Appendix. The reported oracle estimates are based on the best possible $F_{1}$-score over the entire path $\mathcal{P}$. We observe universal improvement of G-StARS over StARS across all tested settings. G-StARS' superior $F_{1}$-score is largely due to drastically improved recall at mildly reduced precision (see Appendix). 
To show the validity of the bounds in B-StARS we also report gap values $\mathrm{gap}_\beta$ and $\mathrm{gap}_\mathrm{b}$ across all settings in Fig.~\ref{fig:gap}. 
Only positive gap values have been measured, thus implying that the StARS bounds have been correct across all tested benchmark problems.

\begin{figure}[h]
\centering
\includegraphics[width=0.99\textwidth]{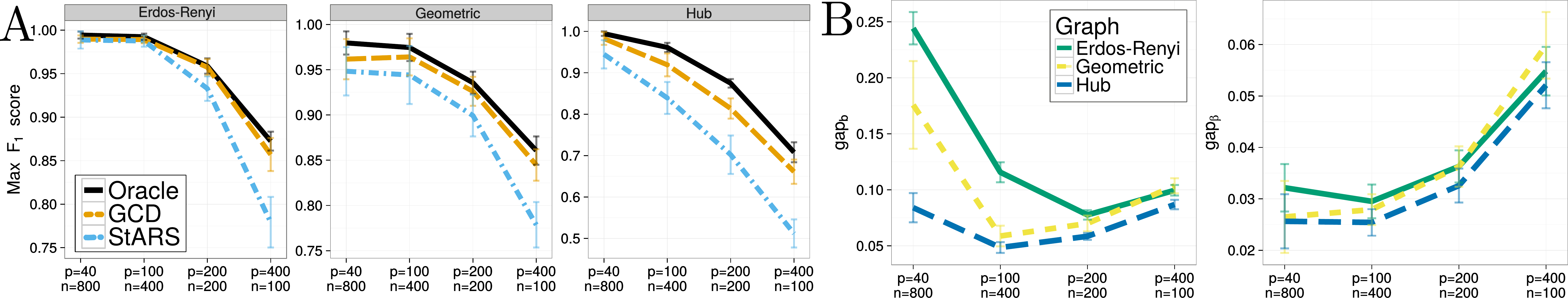}
\caption{A) StARS, G-StARS, and Oracle $F_{1}$-scores \label{fig:f1} B) Observed $\mathrm{gap}_\mathrm{b}$ (left) and $\mathrm{gap}_\beta$ (right).\label{fig:gap}}
\end{figure}
\subsection{Learning large-scale gut microbial interactions}
\label{americangut}
As real-world application we consider learning microbial association networks from population surveys of the American Gut project (see \cite{Kurtz2015a} for pre-processing and data transformation details). The dataset comprises abundances of $p=627$ bacterial taxa across $n=2753$ subjects. We compared run times for StARS and B-StARS in serial and in batch mode over $N=200$ subsamples across $K=100$ values along $\mathcal{P}$. Using B-StARS we found a $44\times$ speed up over StARS (see Tbl.~\ref{table:quicpulsar}) in serial mode on a standard laptop and a $4.5\times$ speed-up in batch mode in a high performance cluster environment. StARS and B-StARS solutions were identical under all scenarios ($\Vert\hat{\Theta}(\lambda_\beta)\Vert_{0}=4612$ at $\lambda_{\beta}=0.194$). On this dataset G-StARS selects $\lambda_{\gamma}=0.127$ leading to a slightly denser graph. We refer to the Appendix for further details and analysis of this test case. 



\section{Conclusions}
In this contribution we have presented a generalization of the state-of-the-art Stability Approach to Regularization Selection (StARS) for graphical models. We have proposed B-StARS, a method that uses lower and upper bounds on the regularization path to give substantial computational speed-up without compromising selection quality and G-StARS, which introduces a novel graphlet stability measure leading to superior graph recovery performance across all tested graph topologies. These generalizations expand the range of problems (and scales) over which Gaussian graphical model inference may be applied and make large-scale graphical model selection a routine task on standard multi-core computers.  


\newcounter{foo}
\renewcommand{\thefoo}{A\arabic{foo}}

\appendix
\counterwithin{figure}{section}
\counterwithin{table}{section}

\newpage
\section*{Appendix: Generalized Stability Approach for Regularized Graphical Models}


\section{Run time and performance for QUIC.}

\begin{table}[h]
  \caption{Erd\"os-Renyi graph with $p=4000$ nodes; \texttt{QUIC} convergence tolerance $\epsilon_Q = 1e-2$; length of path $\mathcal{P}$ is $K=20$.}
  \label{table:quic}
  \centering

\begin{tabular}{ccccccccccc}
\toprule
\multicolumn{2}{c}{{\footnotesize{}Parameters}} &  & \multicolumn{3}{c}{{\footnotesize{}Properties of the solution}} &  &  & \multicolumn{3}{c}{{\footnotesize{}Properties of the solution}}\tabularnewline
\midrule
{\footnotesize{}$\lambda$} & {\footnotesize{}$n$} & {\footnotesize{}time (s)} & {\footnotesize{}$\Vert\hat{\Theta}\Vert_{0}$} & {\footnotesize{}TPR} & {\footnotesize{}FPR} & {\footnotesize{}$n$} & {\footnotesize{}time (s)} & {\footnotesize{}$\Vert\hat{\Theta}\Vert_{0}$} & {\footnotesize{}TPR} & {\footnotesize{}FPR}\tabularnewline
\midrule
{\footnotesize{}$0.08$} & \multirow{5}{*}{{\footnotesize{}$2000$}} & {\footnotesize{}$23$} & {\footnotesize{}$42094$} & {\footnotesize{}$0.81$} & {\footnotesize{}$3\times10^{-4}$} & \multirow{5}{*}{{\footnotesize{}$447$}} & {\footnotesize{}$341$} & {\footnotesize{}$1116870$} & {\footnotesize{}$0.603$} & {\footnotesize{}$0.068$}\tabularnewline
\cmidrule{1-1} \cmidrule{3-6} \cmidrule{8-11} 
{\footnotesize{}$0.05$} &  & {\footnotesize{}$104$} & {\footnotesize{}$442140$} & {\footnotesize{}$0.986$} & {\footnotesize{}$0.025$} &  & {\footnotesize{}$1685$} & {\footnotesize{}$2256934$} & {\footnotesize{}0.702} & {\footnotesize{}$0.139$}\tabularnewline
\cmidrule{1-1} \cline{3-6} \cmidrule{8-11} 
$\mathcal{P}$ &  & {\footnotesize{}$439$} & - & - & - &  & {\footnotesize{}$6056$} & - & - & -\tabularnewline
\bottomrule
\end{tabular}
\end{table}

\section{Illustrating graphlet variability}
\label{subsec:hubex}
To illustrate the multi-modality of graphlet variability $\hat{d}_{N}(\lambda)$ over the regularization path we consider learning a two-component hub graph with $p=40$ nodes (see \citep{Liu2010} and Sec.~5 in main document), containing two hub nodes each having 19 neighbors. For the present example, the weights between hub and peripheral nodes were set to a small value of $-0.117$ where optimal model selection with StARS is challenging even when $n\gg p$. We generated $n=4000$ multivariate normal samples and used \texttt{QUIC} to estimate graphs over $\mathcal{P}$ with $K=100$ equally distributed $\lambda_k$ values for $N=200$ subsamples. 

\begin{figure}[h]
\includegraphics[width=0.99\textwidth]{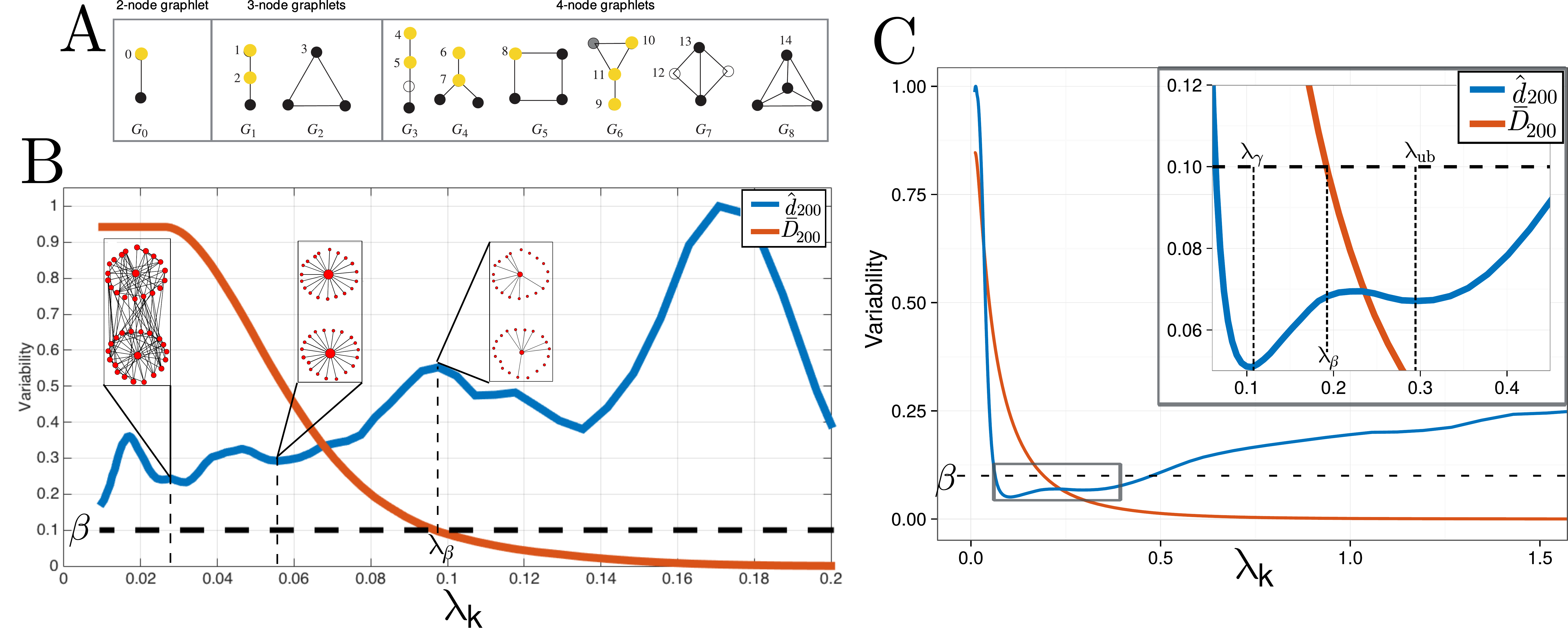}
\caption{A) All graphlets $G_i,i=0,\ldots,8$ of vertex size $\le 4$. The internal vertex numbering refers to all unique 15 orbits. The 11 yellow nodes represent a set of non-redundant orbits used in the Graphlet Correlation Distance (GCD) (adapted from \cite{Przulj2007}).\label{fig:graphlets2} B)Total edge variability $\bar{D}_{200}(\lambda)$ (red curve) and graphlet variability $\hat{d}_{200}(\lambda)$ (blue curve) (scaled to $[0,1]$ for better readability) over the regularization path $\mathcal{P}$ for the two-component hub graph with $p=40$ nodes (see main text for discussion). Same as B) for the American gut data. G-StARS selects the parameters $\lambda_\gamma$ near the global minimum of $\hat{d}_{200}$. Interestingly, $\lambda_\text{ub}$ is also located near a local minimum of the graphlet variability.   
\label{fig:summary1}}
\end{figure}
Figure \ref{fig:summary1}B) shows the traces of $\bar{D}_{200}(\lambda)$ and $\hat{d}_{200}(\lambda)$ over $\mathcal{P}$. While $\bar{D}_{200}(\lambda)$ shows the standard monotonic decrease with increasing $\lambda$, the graphlet variability comprises several local optima. The location of the second highest maximum along $\hat{d}_{200}(\lambda)$ coincides with $\lambda_\beta$  for standard StARS (with $\beta=0.1$). This choice would result in an overly sparse graph. Notably, the true two-component hub graph would be recovered at a local minimum of $\hat{d}_{200}$ near $\lambda_k=0.06$ (at $\bar{D}_{200}\approx 0.5$). Graphs recovered near the lowest local minimum of $\hat{d}_{200}$ (at $\lambda_k\approx 0.03$) would result in a single-component dense graph. In summary, these observations illustrate  that exploring the modes along the graphlet variability curve $\hat{d}_{200}(\lambda)$ can give valuable insights into the evolution of stable graph topologies along the regularization path. 

\section{Graphical model inference from American gut survey data}
\label{americangut}
\begin{figure}[h]
\includegraphics[width=0.99\textwidth]{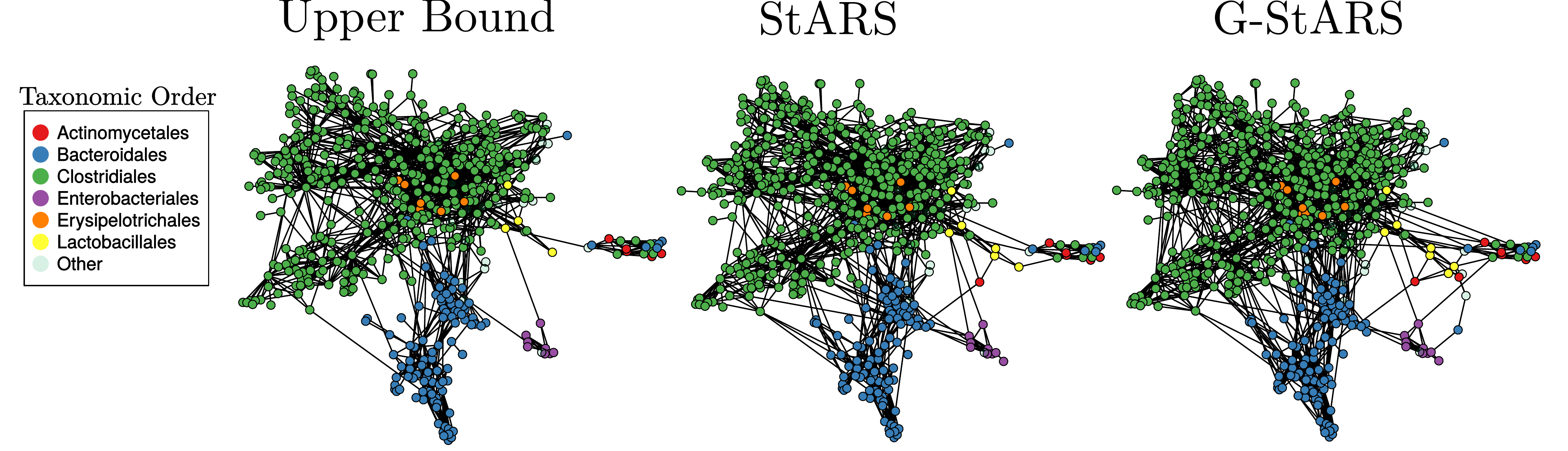}
\caption{Force-directed layout of the inferred gut microbial association graphs at three different regularization parameters (see text for description).
\label{fig:netsummary}}
\end{figure}

The primary data for this application case can be found at \url{https://goo.gl/bW8ZJK}. Figure \ref{fig:summary1}C) shows edge and graphlet variability along the regularization path of the graphical model inference from the American gut data. Highlighted are regularization parameters $\lambda_\text{ub}$ from B-StARS' upper bound, StARS' $\lambda_\beta$, and GStARS' $\lambda_\gamma$. Figure~\ref{fig:netsummary} displays the corresponding gut microbial association networks. 

\begin{figure}[h]
\includegraphics[width=0.99\textwidth]{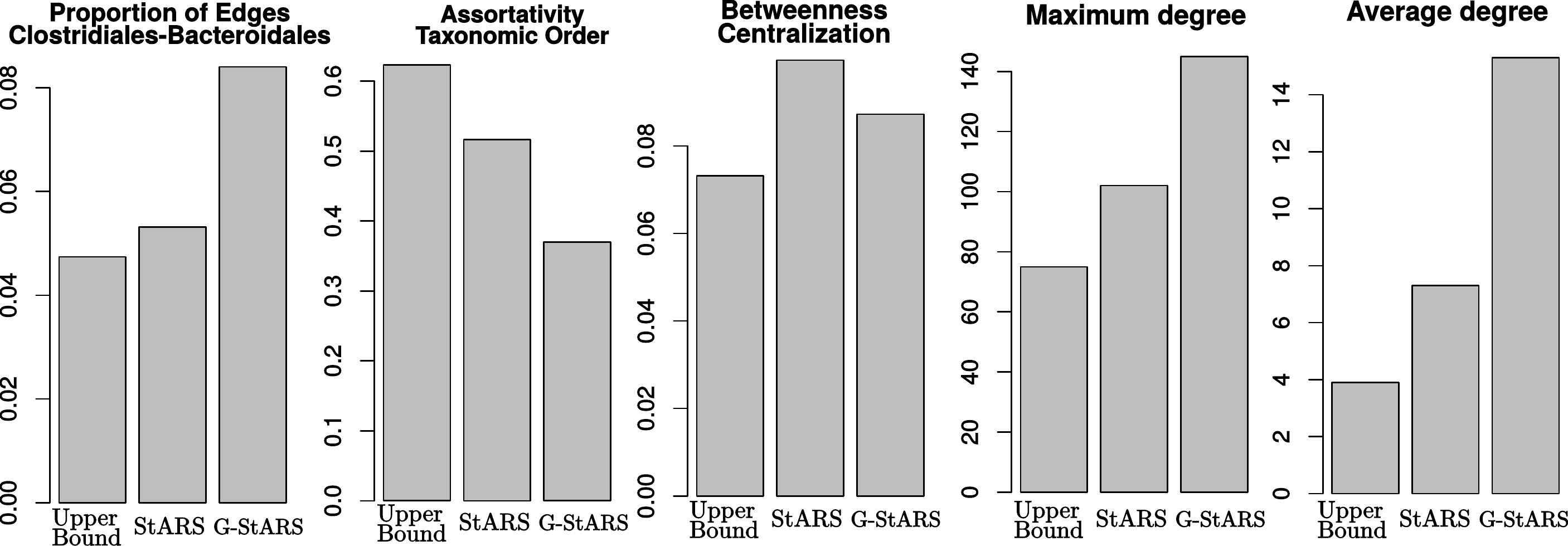}
\caption{Summary statistics of the topology from inferred gut microbial association graphs.
\label{fig:summarystats}}
\end{figure}

Figure~\ref{fig:summarystats} displays different global graph features of these networks. While no verified gold standard for microbe-microbe associations are available for these data, we found a higher proportion of edges between Clostridiales and Bacteroidales nodes (see first bar plot in Fig~\ref{fig:summarystats}) in the graph selected by G-StARS. This is consistent with recent experimental and statistical evidence of negative associations between members of these orders \citep{Ramanan2016}.

\bibliographystyle{apalike}
\bibliography{gstars}
\end{document}